\documentclass[aps,pra,english,showpacs,twocolumn]{revtex4}
\usepackage[english]{babel}
\usepackage[ansinew]{inputenc}
\usepackage{amsfonts}
\usepackage{latexsym}
\usepackage{amsmath}
\usepackage{graphicx}

\begin{document}

\title{Non-Gaussian states from continuous-wave Gaussian light sources}

\author{Klaus Mølmer}
\affiliation{Danish National Research Foundation Center for Quantum Optics\\Department of
Physics and Astronomy\\University of Aarhus\\DK 8000 Aarhus C, Denmark}

\date{\today}

\begin{abstract}
We present a general analysis of the state obtained by subjecting the output from a continuous-wave
(cw) Gaussian field to non-Gaussian measurements. The generic multimode state of cw Gaussian fields is
characterized by an infinite dimensional covariance matrix involving the noise correlations of the
source. Our theory extracts the information relevant for detection within specific temporal output
modes from these correlation functions . The formalism is applied to schemes for production of
non-classical light states from a squeezed beam of light.
\end{abstract}

\pacs{03.65.Wj; 03.67.-a; 42.50.Dv}  \maketitle

\section{Introduction}

Quantum computing and communication, and quantum metrology with atoms and light are topics of high
current interest offering on the one hand proposals for new revolutionary technologies and addressing
on the other hand fundamental questions about the properties of quantum systems. Light is an ideal
carrier of information over long distances, and it is an ideal physical probe for the properties of
matter. Numerous proposals have been made for the use of light for quantum information purposes and
both theory and experiments have advanced to the quest for and application of specially designed
quantum states of light. The observation that coherent states of light suffice for quantum
cryptography \cite{coh-crypt} has not removed the interest in single-photon and other non-classical
states, since it has been proven that the entire family of Gaussian states (to which the coherent
states belong) cannot be distilled by Gaussian operations \cite{no-distill}, and since, e.g., the
implementation of qubits for digital quantum computing also necessarily involves non-Gaussian states
\cite{klm}.

Non-Gaussian states can be produced by single photon emitters such as atoms or ions in cavities
\cite{rempe,lange}, or N-V centers \cite{nv-centers} (for a recent review of single-photon emitters
see \cite{single-review}), but they may also result from post selecting the output of Gaussian light
sources after non-Gaussian measurements have been carried out on part of the field \cite{dakna}.
Thermal, coherent and squeezed light are all examples of Gaussian states, and prompted by recent
experimental and theoretical work on single photon production from a cw squeezed light field
\cite{wenger,neergaard,grosshans,sasaki,kim,laflamme}, we shall in this paper develop a theory for the
state of a specific output mode, obtained from a cw Gaussian state light after an arbitrary
measurement has been performed on a designated trigger mode. This analysis in fact consists in two
straightforward steps:

\noindent 1. From the general multimode Gaussian state, fully characterized by the correlation
functions of the light beam, we derive the covariance matrix and corresponding Gaussian Wigner
function for the state of the (trigger+output) two mode system prior to detection.

\noindent 2. The two-mode Wigner function has two complex (four real) arguments, and the effect of the
measurement on the trigger mode is described by a (generalized) projection, which amounts to a
specific operation on the Wigner function followed by an integration over the trigger mode variables.

These steps, which can both be carried out analytically for several relevant cases, produce a Wigner
function for the quantum state of the output mode which provides all available information about the
state.

In Sec.II, we show explicity how to extract the two-mode Gaussian state from the noise correlations of
the Gaussian field light source. In Sec. III, we discuss the state reduction and the output field
resulting from different kinds of photon counting measurement on the trigger mode. In Sec. IV, we
specialize to the interesting case of single photon production, and we present a concrete analysis for
the squeezed output from an OPO and  we summarize the technical implementation of various elements and
phenomena (beam splitters, losses, inefficient detectors, dark counts, ...). Sec. V concludes the
paper.

\section{Gaussian states - reduction to two modes}

In a generic description of the experiments that we wish to analyze, a cw Gaussian light field is
split and possibly filtered by a sequence of linear optical devices, and one particular field mode (in
our example: the field at a specific instant of time) is registered by a measurement device, see Fig.
1. Conditioned on the measurement output, another component of the beam, and in particular the field
pertaining to a single designated output mode, is emitted in a non-Gaussian state. In the case of a
low flux from an OPO, the total field contains vacuum and a weakly populated two-photon component, and
a single detector click, stemming from the two-photon component, suggests that a one photon state is
produced. At higher flux, the squeezed output is a superposition of even photon number states, and a
single photon subtraction returns a superposition of only odd number states, which is a non-classical
state with some Schr\"odinger Cat character. This paper addresses how good single photon states and
how much non-classical character can be obtained in a realistic physical set-up.

\begin{figure}
  \includegraphics[width=7.5cm]{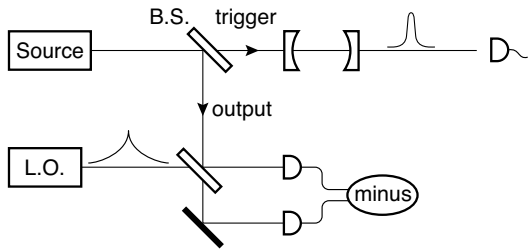}\\
  \caption{Experimental set-up for production of a non-Gaussian state from a Gaussian cw source of light.
  The field is split into trigger and output components, and after filtering (optional), a non-Gaussian
  photon-counting detection takes place on the trigger mode. Conditioned on the trigger signal, the output beam
  may be in a non-classical state, and by use of a temporally modulated local oscillator field, a specific single
  mode in the output beam can be extracted for homodyne detection.}\label{Fig1}
\end{figure}
Our analysis proceeds by noting that prior to the count event, we have a system with two relevant
field modes:

\noindent \emph{mode 1}: the temporal  mode on which the trigger detection takes place. We assume an
instant click event at time $t_c$, but the field may well be frequency filtered prior to detection, so
that the field registered is actually an integral over time of the field that left the original light
source.

\noindent \emph{mode 2}: the output mode, populating a quantum state that we shall characterize fully
by its phase space Wigner function. The Wigner function is experimentally available by homodyne
detection and quantum state tomography \cite{tomography}, where the homodyne detection can be carried
out with a time dependent local oscillator field, which will extract the desired output mode function.
For our concrete application we expect the optimum mode function to be localized around the time of
the click event on mode 1, and that it has a temporal shape reflecting the band width of the squeezed
light source. See Fig.2.

\begin{figure}
  \includegraphics[width=7.5cm]{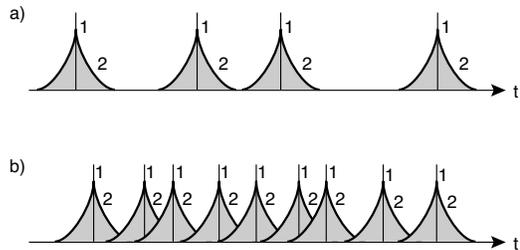}\\
  \caption{Schematic representation of the trigger (1) and output (2)  signals. The OPO produces photons in pairs inside a cavity,
  and when a discrete trigger photon is
  detected (vertical lines in the figure), we expect to have a second photon in the output beam, temporally separated from the trigger event according to a
  mode function (shaded function in the figure) with a width related to the
  squeezing band width of the source (part a). If the squeezed light source is too bright, the click
  events occur more often, and the mode functions of the output single photons states start to
  overlap in an incoherent manner (part b).}
\end{figure}

Let $\hat{a}(t)$ and $\hat{a}^\dagger(t)$ denote the annihilation and creation operators of the source
field. The input to our analysis is the noise correlations represented by the correlation functions
$\langle \hat{a}(t)\hat{a}(t')\rangle$ and $\langle \hat{a}^\dagger(t)\hat{a}(t')\rangle$. We do not
have to restrict the analysis to this case, but we note that for a stationary source, these
correlation functions only depend on the difference between time arguments $(t-t')$. The most general
formulation of the trigger+output modes and their properties is obtained by designating field mode
operators,
\begin{equation} \label{generalmode}
\hat{a}_i = \int f_i(t') \hat{a}(t')dt' + \xi_i \hat{b}_{i,vac}.
\end{equation}
where the functions $f_i(t)$ with $i=1,2$ denote the mode functions of the trigger and output modes
respectively, taking into account completely arbitrary manipulations of the field states by linear
filters and beam splitters, and detection in arbitrarily shaped mode functions. The vacuum fields
$\hat{b}_{i,vac}$ are needed for unitarity, but their effects are fully accounted for by the source
terms.

With explicit expressions for the functions in (\ref{generalmode}) and for the source correlations, we
can compute all second moments $\langle \hat{a}_i\hat{a}_j\rangle$, $\langle\hat{a}_i^\dagger
\hat{a}_j\rangle$ of the trigger-output modes prior to the trigger detection. The cw multi-mode field
state is Gaussian, and when the state is restricted to only two modes (all others are traced out) the
Gaussian property remains. This implies that the two-mode Wigner function of the field modes 1 and 2
is a Gaussian function of four real quadrature variables $(y_1,y_2,y_3,y_4)=(x_1,p_1,x_2,p_2)$, fully
characterized by a covariance matrix with elements $V_{ij}\equiv<y_i y_j>+<y_j y_i>$, expressed
explicitly in terms of $\langle\hat{a}_i\hat{a}_j\rangle$, $\langle\hat{a}_i^\dagger\hat{a}_j\rangle$.
\begin{equation}\label{wigner}
W_V({\bf y}) = \frac{1}{\pi^2\sqrt{\textrm{det}(V)}} \exp(-{\bf y}^T V^{-1}{\bf y}),
\end{equation}
where ${\bf y}$ is a column vector, and $\cdot^T$ denotes the transpose. (We assume without loss of
generality, that there are no mean field components and hence no displacement of the state to take
care of).

\section{Non-Gaussian states}

The Wigner function (\ref{wigner}) is so simple, that it constitutes a natural starting point for
analysis of the state of the output mode 2, conditioned on various possible measurements performed on
mode 1. Homodyne or heterodyne measurements are readily accounted for by elementary transformations of
the covariance matrix, returning a Gaussian state of the output field with a mean amplitude that
depends on the trigger measurement outcome, and with a covariance matrix that is independent of the
actual outcome \cite{eisert,larsklaus}. The purpose of the present work is to go beyond the Gaussian
states, and for this purpose more general transformations of the Wigner function (\ref{wigner}) must
be allowed. Such transformations follow if the trigger mode is made subject to a measurement that does
not preserve the Gaussian character, for example a photon counting experiment, and we shall analyze
three variants of photocounting measurements:

\subsection{Perfect number detector}

A perfect photon number detector will yield an integer eigenvalue, $n=0,1, ...$, when applied to the
trigger mode. Such a detection event is accompanied by a projection of the field state on the
corresponding trigger mode eigenstate with the Wigner function $W_n(x_1,p_1)$, which for the remaining
output mode implies a change into a state described by the conditioned Wigner function
\begin{equation}\label{onephoton}
W_{c,n}(x_2,p_2) = {\cal N}_n \int dx_1 dp_1 W_V(x_1,p_1,x_2,p_2) W_n(x_1,p_1),
\end{equation}
where the normalization constants ${\cal N}_n$ are inversely proportional to the probabilities for the
different number state outputs. The most interesting case is probably the one of a detection of a
single photon with
\begin{equation}
W_{n=1}(x_1,p_1)=\pi^{-1}(2 x_1^2+2 p_1^2-1)\exp(-x_1^2-p_1^2),
\end{equation}
and we observe that the integral (\ref{onephoton}) only involves a Gaussian and a product of quadratic
terms and a Gaussian. It can hence be carried out analytically, the result being itself a product of a
Gaussian and a second order poynomial in the variables $x_2,p_2$. Kim et al \cite{kim} have studied
the negativity of the Wigner function at the origin as a suitable measure of the non-classicality of
the output of such a protocol. Their study dealt only with a physical two-mode system, but they
assumed a sufficiently general covariance matrix $V$ that they describe also our general cw setting,
where $V$ is the result of an integration over the cw field correlation functions and arbitrary
trigger and output mode functions. Their analytical expression for the conditioned Wigner function
$W_{c,n=1}(x_2,p_2)$ and in particular for its possible negativity at the origin, may be helpful in
the search for optimum trigger and output mode functions in such an experiment. Dakna et al
\cite{dakna} have studied the case of a pure single-mode squeezed state, which after beam splitting is
a special case of the Gaussian two-mode states, and they have presented various results including
output mode Wigner functions conditioned on the detection of different number states of the trigger
mode. With Eqs.(\ref{wigner},\ref{onephoton}), and the appropriate covariance matrix, determined from
the source correlation functions, we are able to generalize these results to search for Schr\"odinger
cat-like states in cw experiments.

\subsection{"On/off" detector}

The so-called on/off detector, is able to distinguish between a vanishing and a non-vanishing photon
number in a field mode, but it is unable to resolve photon numbers larger than unity. The interesting
"on" output of such a detector, projects the system on the space spanned by all non-vanishing photon
number states, equivalent to removal of the vacuum component, and the resulting Wigner function of the
output mode is
\begin{eqnarray}\label{on}
W_{c,n}(x_2,p_2) = {\cal N}_{on} \int dx_1 dp_1 W_V(x_1,p_1,x_2,p_2)\nonumber \\ (1-W_{n=0}(x_1,p_1)),
\end{eqnarray}
where $W_{n=0}(x_1,p_1)=\pi^{-1}\exp(-x_1^2-p_1^2)$, and the integral is readily evaluated for any
covariance matrix $V$. The result is readily seen to be the difference between two Gaussian functions
in the variables $x_2,p_2$. This detector model was recently applied by Sasaki and Suzuki in a
theoretical analysis \cite{sasaki} of the cw OPO output field. Their approach did not reduce the
system first to the two relevant modes, but used instead a mode expansion on a complete basis of
prolate spheroidal wave functions and suitable generating functions for the number distributions
studied also by Zhu and Caves \cite{zhu}. They do, however, also obtain Wigner functions expressed
formally as the difference between two Gaussian functions with different pre-factors and widths. Given
that both the covariance matrix $V$ and the integral (\ref{on}) can be evaluated analytically, we
advocate that our approach provides a more straightforward analysis of the "on/off" detector.

\subsection{Click event}

The most natural and realistic detector model, is the one of usual photo detection theory, in which
the trigger mode is incident on a physical material, from which a bound electron is excited into the
continuum and subsequently detected. In this scenario, it is the electron that is being detected and
made subject to a projection into the continuum state space, and the back action on the field state is
given by the field annihilation operator \cite{glauber}. In a density matrix formulation this amount
to the transformation $\rho \rightarrow \hat{a}_1
\rho\hat{a}_1^\dagger/\textrm{Tr}(\hat{a}_1^\dagger\hat{a}_1 \rho)$, and using the operator
correspondence \cite{gardinerbook} between application of the operator
$(\hat{a}_1=(\hat{x}_1+i\hat{p})/\sqrt{2}$ and multiplying and differentiating the Wigner function
with respect to its arguments $x_1,p_1$, we find the non-Gaussian two-mode Wigner function of the
field after the click of the detector. The reduced state of the output field is at this point obtained
by tracing over the trigger mode field, i.e., by integrating the Wigner function with respect to the
$x_1,p_1$ arguments yielding the final expression,
\begin{eqnarray}\label{click}
W_{click}(x_2,p_2) = {\cal N}_{click} \int dx_1 dp_1\nonumber \\ \frac{1}{2}\left(
x_1^2+p_1^2+\frac{1}{4}(\frac{\partial^2}{\partial x_1^2}+\frac{\partial^2}{\partial
p_1^2})+x_1\frac{\partial}{\partial x_1} + p_1\frac{\partial}{\partial p_1} + 1\right)\nonumber \\
W_V(x_1,p_1,x_2,p_2).
\end{eqnarray}
This is a lengthy expression, but the appearance of terms up to only second order together with the
Gaussian function ensures that the result, as in the case of a one-photon detection, is of the form of
a Gaussian function multiplied with a second order polynomial with no first order terms in $x_2,p_2$.
In practice, the expressions become unwieldy, but the analytical toolboxes in Mathematica, Matlab, or
similar solvers readily deal with the expressions.

\section{Results for a realistic experiment}

In this section, we consider an optical parametric oscillator (OPO) with mirrors with leakage rates
$\gamma_1,\gamma_2$ and a nonlinear gain $\varepsilon$. The output field from this device has the
correlation functions \cite{gardinerbook}
\begin{eqnarray} \label{correl}
\langle \hat{a}(t)\hat{a}(t')=\frac{\gamma_1}{\gamma_1+\gamma_2}\frac{\lambda^2-\mu^2}{4}
(\frac{e^{-\mu |t-t'|}}{2\mu}+\frac{e^{-\lambda |t-t'|}}{2\lambda})\nonumber \\
\langle \hat{a}^\dagger(t)\hat{a}(t')=\frac{\gamma_1}{\gamma_1+\gamma_2}\frac{\lambda^2-\mu^2}{4}
(\frac{e^{-\mu |t-t'|}}{2\mu}-\frac{e^{-\lambda |t-t'|}}{2\lambda})
\end{eqnarray}
where $\lambda=\frac{1}{2}(\gamma_1+\gamma_2)+\varepsilon$, $\mu
=\frac{1}{2}(\gamma_1+\gamma_2)-\varepsilon$, and where the time dependent field operators are
normalized according to delta-correlated commutators $[\hat{a}(t),\hat{a}^\dagger(t')]=\delta(t-t')$.

The trigger mode 1 and output mode 2 are single modes of the field, which are in the most general case
written as in Eq.(\ref{generalmode}) with contributions from the cw source and from vacuum noise
sources. Let us comment on the explicit form of these expressions: In our example we imagine that part
of the OPO beam is transmitted with amplitude $\tau$ through a beam splitter and the time dependent
annihilation operator for the transmitted field can be written
\begin{equation}
\hat{a}_\tau(t)=\tau\hat{a}(t) + \rho \hat{b}_{vac}(t), \label{a10}
\end{equation}
where $\rho \hat{b}_{vac}(t)$ is a vacuum noise contribution, needed for unitarity, but causing no
complications in the following. Field transmission losses and a finite detector efficiency for the
trigger mode can be modelled by the effect of yet another beam splitter that transmits only a certain
fraction of the field to a perfect counter and adds extra vacuum contributions, i.e., by a simple
reduction of the factor $\tau$ in Eq.(\ref{a10}) and by replacement of $\rho$ by a correspondingly
larger number.

Further manipulation of the field prior to detection may be relevant, for example a frequency filter
may be installed to avoid contributions from other modes of the OPO, not accounted for by
Eq.(\ref{correl}), see for example \cite{niu}. A simple frequency filter of width $\gamma$  has an
exponential temporal transmission function, and the time dependent annihilation operator for the
filtered field can be written
\begin{equation}
\hat{a}_\tau(t)=\tau\gamma\int_{-\infty}^t exp(-\gamma (t-t')) \hat{a}(t') dt' + \beta_1
\hat{b}_{1,vac}(t), \label{a1}
\end{equation}
where $\beta_1 \hat{b}_{1,vac}(t)$ is again a short hand for the vacuum noise contribution.

Eq.(\ref{a1}) still represents a cw field, but we now assume that this field is being observed by some
photon counting device, and that we can define sharply peaked mode functions centered at different
times with corresponding annihilation and creation operators with standard single mode commutators.
Three different counting operations were described in the previous section, and in a typical
experimental implementation of the \emph{click detection} the field will be monitored continuously in
time, and eventually, after an interval of no clicks, the counting device reports a click at time
$t_c$. We shall hence focus our attention on the correlations between the particular discrete trigger
mode where this happens and the remaining output from the light source. We thus define the single mode
operator $\hat{a}_1=\int h_{t_c}(t)\hat{a}_\tau(t)dt$, where $h_{t_c}(t)$ is a normalized sharply
peaked function around $t=t_c$. If $h_{t_c}(t)$ is much narrower than the temporal correlations of the
source field, its actual shape is insignificant, and in a numerical analysis, one may discretize time
in intervals of length $dt$ and simply take $h_{t_c}(t)=1/\sqrt{dt}$ on the time interval including
$t_c$. We note that $\hat{a}_1$ is now fully determined as a specific integral over the source output
field annihilation operator and vacuum contributions, defining thus precisely the function $f_1(t')$
in Eq.(\ref{generalmode}).

The whole purpose of the physical setup is to produce a useful non-Gausssian output state, and the
output field, which is the component reflected by the first beam splitter in Fig.1, should now be made
subject to investigation, conditioned on the click event on the trigger mode. It should be expected,
that our non-classical output state predominantly occupies a temporal mode of the field that left the
OPO cavity at the same time as the field component that caused the detector click on the trigger mode,
and we will also expect that the mode has a temporal width related to the bandwidth of the OPO cavity
as represented by the parameters $\lambda$ and $\mu$ in Eq.(\ref{correl}). Using the correlation
functions (\ref{correl}), one may indeed determine the conditioned coherence function $\langle
\hat{a}^\dagger(t_c)\hat{a}^\dagger(t)\hat{a}(t')\hat{a}(t_c)\rangle$, which in the spirit of standard
photo detection theory describes the coherence properties of the field, conditioned on a click event
at time $t_c$. This function is readily calculated for the specific correlation functions of the OPO,
and for a low flux (small $\varepsilon$) one indeed finds that this function factors into a product
$u^*(t)u(t')$ with a mode function $u(t)$ falling off symmetrically and exponentially as function of
$|t-t_c|$, as illustrated in  Fig.2a. For larger fluxes, the function no longer factors, pointing to
the multi mode character of the field, which we can understand as contributions to the field which are
correlated with the not so distant click events at other times, as illustrated in Fig. 2b.

Our formalism permits the definition of mode functions $u(t)$ of arbitrary shape, and in a homodyne
detection set-up, the extraction of a the field quadrature distribution  of a particular mode is
implemented by the corresponding temporal modulation of the local oscillator field. We can model
losses and imperfect detectors by introducing the effects of an equivalent beam splitter, and hence
the single output field mode 2 is also described by Eq.(\ref{generalmode}) with an explicit expression
given for $f_2(t')$.

With explicit expressions for the functions $f_i(t')$ and the correlations (\ref{correl}), we
determine by straightforward integration the correlations between the two discrete modes 1 and 2, and
subsequently their covariance matrix $V$.  Losses and possible incoherent stray field components which
may lead to false counts can also be included effectively in the formalism by a modification of the
covariance matrix $V \rightarrow LVL+N$, where a diagonal matrix
$L=\textrm{diag}(\sqrt{1-\eta_1},\sqrt{1-\eta_1},\sqrt{1-\eta_2},\sqrt{1-\eta_2})$ describes the
transmission (or detection) efficiencies $1-\eta_i$ for the modes, and where
$N=\textrm{diag}(\eta_1+\xi_1, \eta_1+\xi_1, \eta_2+\xi_2, \eta_2+\xi_2)$ describes the corresponding
vacuum ($\eta_i$) and additional ($\xi_i$) noise contributions, see for example Ref.\cite{larsklaus}.

We shall now show a few examples of Wigner functions for output fields, obtained for different
parameters. In the following we show the result for an OPO  with one perfectly reflecting mirror
$\gamma_2=0$ and we shall provide the coupling strength of the squeezing Hamiltonian $\varepsilon$,
and the rates and temporal variation, characterizing the mode functions $f_i(t)$, in unit of
$\gamma_1$.

We take first a weakly excited OPO with $\varepsilon=0.01\gamma_1$. We assume the triggering
beamsplitter to have a transmission of 1 \%, and we frequency filter the trigger field with a width of
$\gamma=5 \gamma_1$ (has only little effect on the field) before the detection in a narrow ($\gamma_1
dt=0.02$) time window. For simplicity we assume an output mode function of the form $u(t) \propto
\exp(-\alpha|t-t_c|$, with $\alpha=\gamma_1/2$. No losses are assumed. The Wignerfunction is shown in
Fig.3. It looks very much like the Wigner function of a one-photon number state, and we indeed compute
the overlap with the one photon state to be as high as 98.82 \%. In the lower panel of Fig.3. we show
the results obtained if the output light mode suffers a 25 \% loss. The overlap with the one photon
state is now reduced to 74.14 \%. The Wigner function is negative at the origin, with a value of
-0.3116 (no losses) and -0.154 (25 \% loss), compared to the ideal value $-1/\pi=-0.3183$.
\begin{figure}
  \includegraphics[width=7.5cm]{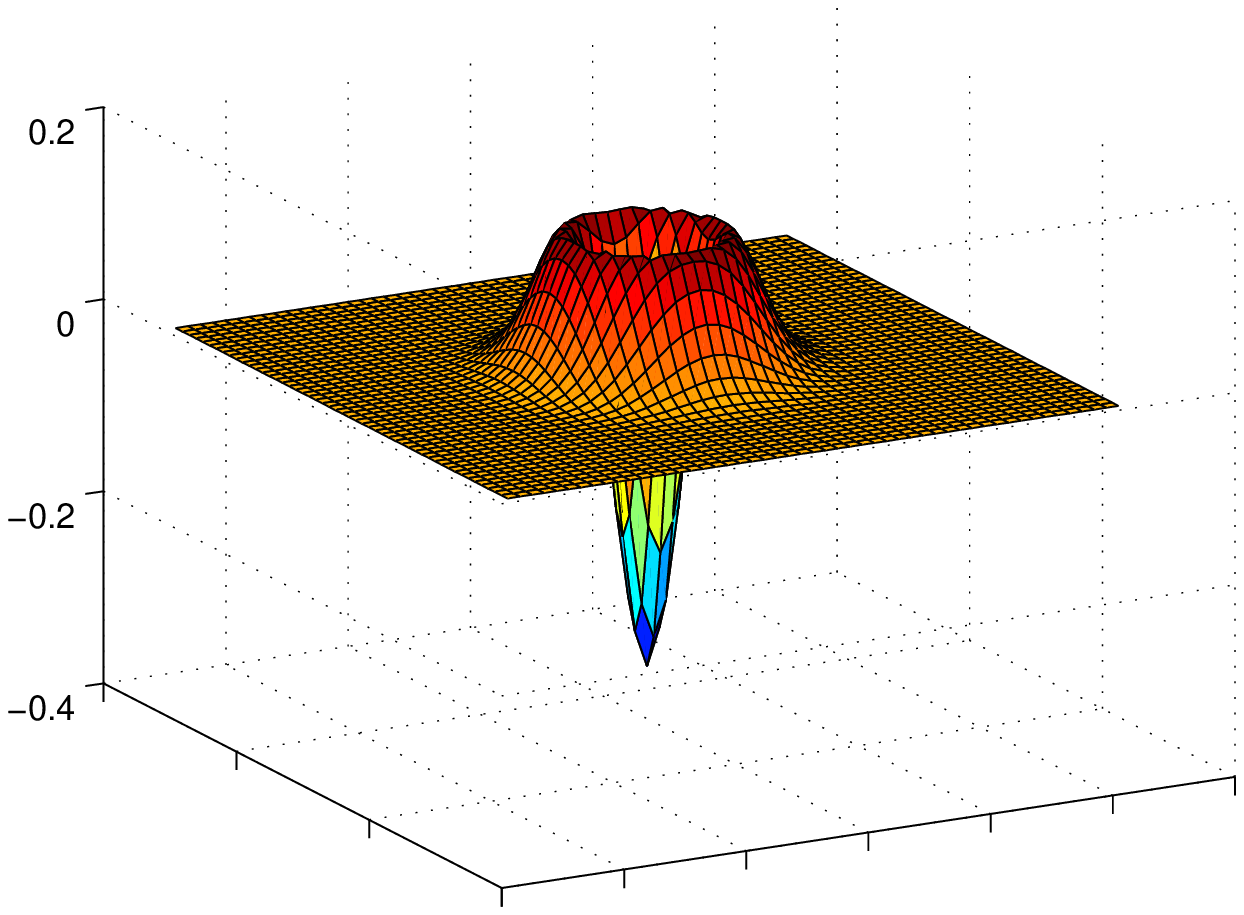}\\
  \includegraphics[width=7.5cm]{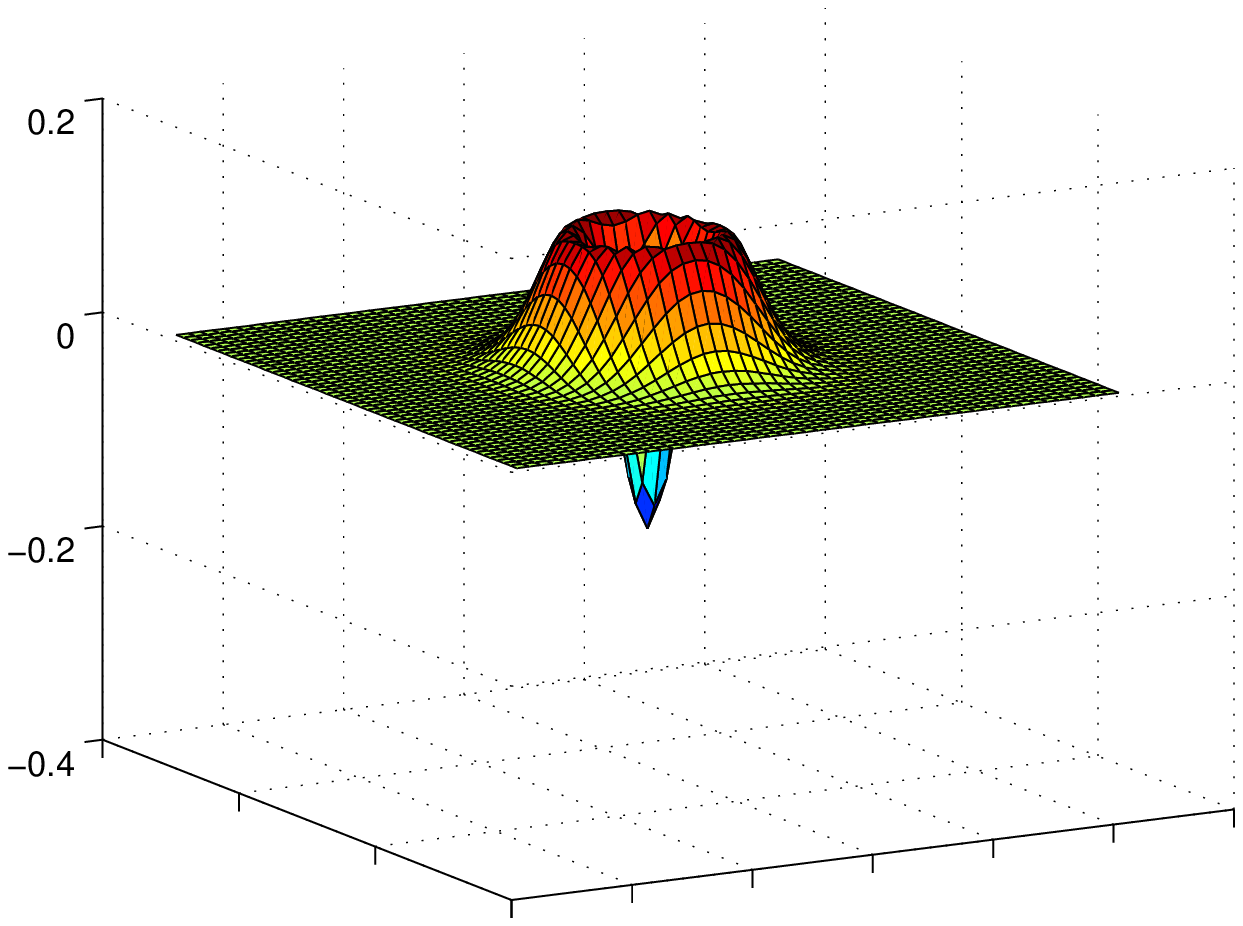}\\
  \caption{Upper panel: Wigner function for the single mode output field with mode function $\exp(-\gamma_1 |t-t_c|/2)$ conditioned on a
  detector click in the trigger mode. The field comes from an OPO with $\varepsilon=0.01\gamma_1,\ \gamma_2=0$, and only 1 \% of the light
  is transmitted to the trigger detector. The lower panel shows the Wigner function obtained with the same parameters as in the upper
  panel but with 25 \% loss in  the output field.}\label{Fig3}
\end{figure}

At higher power levels, the OPO does not produce a one-photon state in coincidence with the click
event on the trigger mode, but a higher excited non-classical state. In Fig. 4, we show Wigner
functions obtained with the same parameters as in Fig. 3, but with $\varepsilon=0.2\gamma_1$. These
figures do not look like one-photon state Wigner functions, but they would be very attractive to
produce and study for their non-classical behavior, and in particular for their double-peaked
Schr\"odinger cat-like character and for the negative values attained at the origin. Without losses,
we find the value of the Wigner function at the origin to be $-0.2499$ in the upper panel and
$-0.0889$ in the lower panel.

\begin{figure}
  \includegraphics[width=7.5cm]{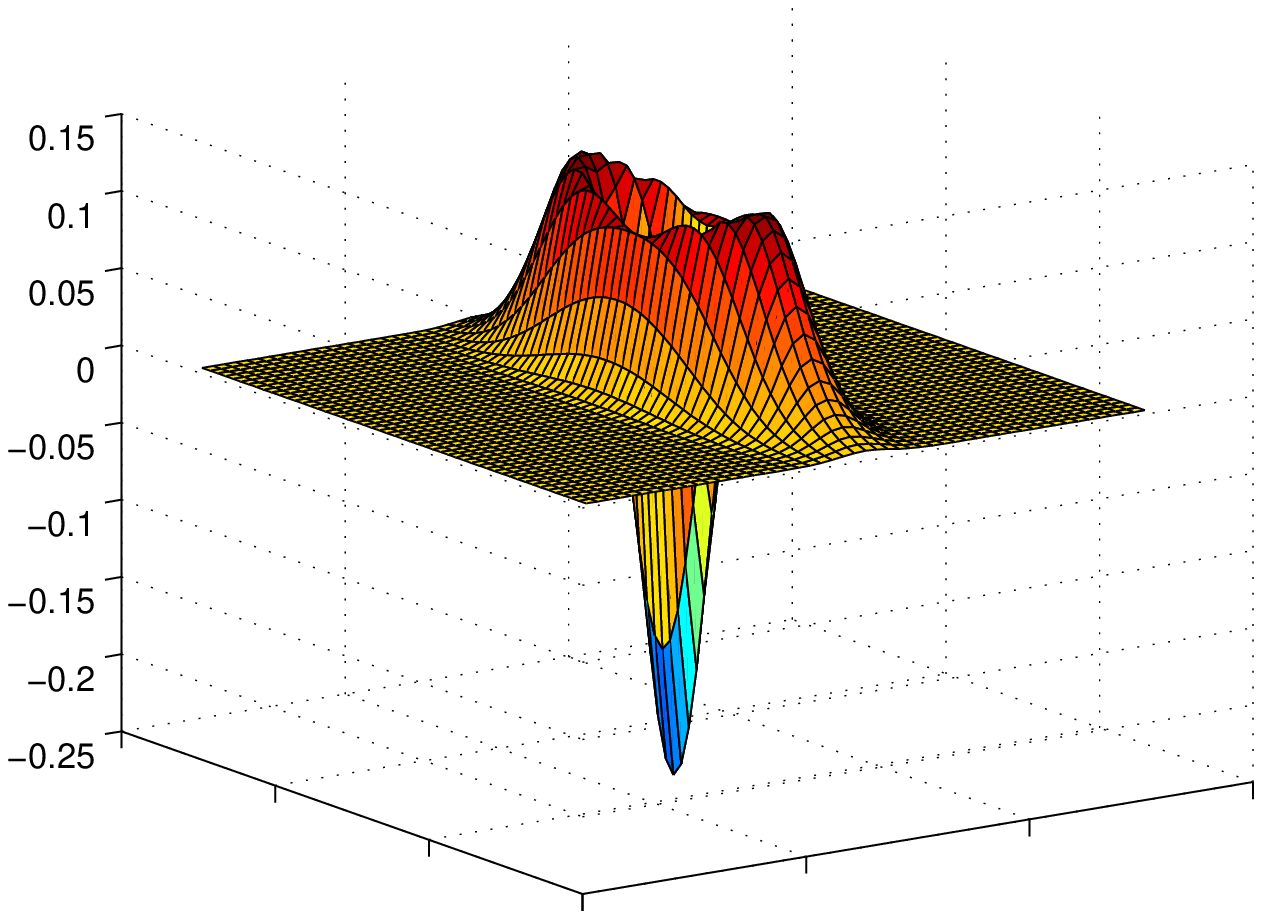}\\
  \includegraphics[width=7.5cm]{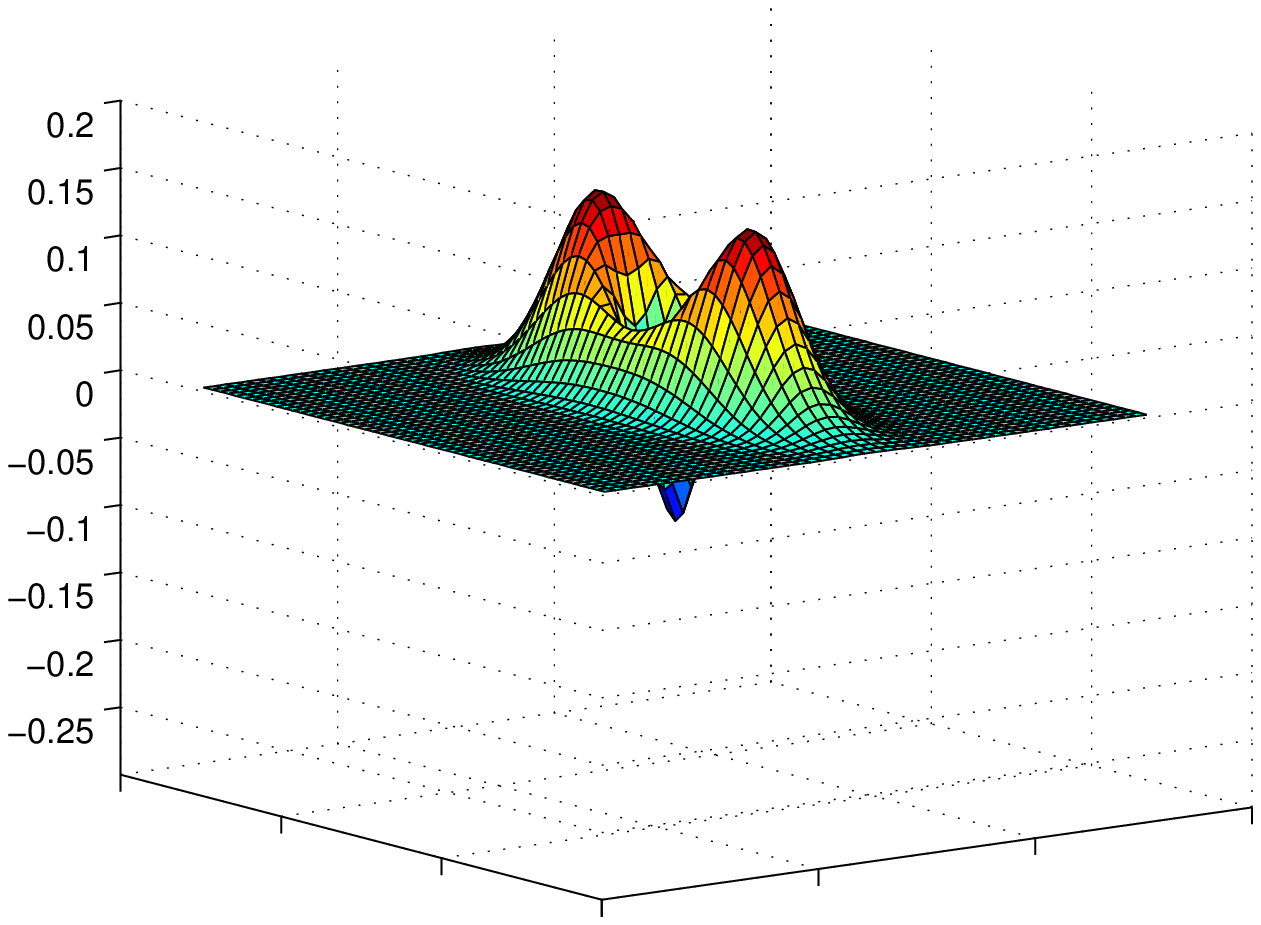}\\
  \caption{Upper panel: Wigner function for the single mode output field with mode function $\exp(-\gamma_1 |t-t_c|/2)$ conditioned on a
  detector click in the trigger mode. The field comes from an OPO with $\varepsilon=0.2\gamma_1,\ \gamma_2=0$, and only 1 \% of the light
  is transmitted to the trigger detector. The lower panel shows the Wigner function obtained with the same parameters as in the upper
  panel but with 25 \% loss in the output field.}\label{Fig4}
\end{figure}

We have performed the calculations with the same parameters as in Figs.4, but allowing for a simple
variation of the output mode function applied in the analysis. If we choose a modified value of
$\alpha=0.337$ instead of $0.5$ for the decay rate of the exponential output mode function
$\exp(-\alpha|t-t_c|)$, the negativity at the origin gets slightly deeper (-0.2618) than in the upper
panel of Fig.4.There is a host of parameters that can be varied more or less systematically, and if
one wishes to produce a specific non-classical state with high fidelity, the present analysis is a
good starting point. The natural next step is then to take the mode functions $f_i(t)$ as variational
functions, and to optimize the quantity of interest (one-photon overlap, negativity at the origin,
cat-like behavior, ...) with respect to these functions.

\section{Discussion}

In conclusion we have established a formalism and we have presented results for the output of cw
Gaussian light sources subject to non-Gaussian discrete measurement events on part of the field. The
starting point of the analysis is the covariance matrix of the full Gaussian multi-mode field, given
explicitly by the continuous time correlation functions (\ref{correl}). The relevant two modes are
described by a reduced quantum state, where one should formally trace over all other unobserved modes.
The infinite dimensional covariance matrix is then reduced to a covariance matrix for the two modes of
interest, and rather than taking the trace, we identify this matrix by calculating explicitly the
covariance matrix elements. The Gaussian character of the two-mode state is broken by the measurement,
but at this stage, the system is so simple that any realistic trigger mode detection scheme is readily
implemented as a formal operation on the two-mode Wigner function of the joint system. This is
illustrated by the action of three different non-Gaussian counting operations, which can all be
handled analytically due to the simplicity of Gaussian integrals of polynomials. A numerical
two-dimensional integral can of course be applied, if need be.

We note that if an experiment is continuously not showing any counting events until time $t_c$, these
null measurements also have implications on the output state, and one might question the use of the
steady state two-time correlation functions in the determination of the output field. In order to rely
on steady state correlation functions, one precisely has to run the experiment continuously in time
and note that click events will occur after sequences of no-click intervals of varying duration, and
that averaging over all click events corresponds to an averaging over all such possible histories,
which, in turn, is equivalent to the steady state average of the unprobed system. This justifies the
theoretical treatment, but it also suggests, that one may carry out experiments, where the output
field is conditioned on a more detailed requirement of the trigger history, for example, that no
clicks have occurred for a certain time before $t_c$. This scenario will be of relevance if the
trigger mode has a high photon detection efficiency (otherwise we automatically ignore many photons
and hence de facto work under average steady state conditions). Since the no-click event is a
projection on the vacuum state and is hence a Gaussian operation, we may in fact provide a simple
theory also for this scenario.

A useful feature of the Gaussian states is the ease at which the number of modes can be increased
without rendering the theoretical treatment untractable. This implies that the above analysis readily
allows generalization to situations where fields from several Gaussian cw and pulsed sources are
included. We imagine in particular, that a recent proposal \cite{laflamme} for production of
one-photon states with no pollution from higher number state components involving two relatively
intense OPO outputs and a coincidence of two single photon detection events, can be studied under true
cw conditions. Such an analysis would imply a second step away from Gaussian states, and it remains to
be seen which tools will be optimum for the efficient handling of this progressively more complicated
problem.

Despite the fact that the quantum theory of electromagnetic radiation has been well established for a
very long time and that all aspects of light generation and detection are quantum optics text book
material, a "gap" remains between the description of light as a field and as composed of photons. This
gap is not due to the wave-particle duality in quantum theory, but is of a much more pragmatic origin,
namely that the expansion of fields in terms of number states is sensible and useful only for a single
or few modes, whereas cw fields with infinitely many modes are more conveniently described in terms of
the properties of field operators, labelled by continuous mode arguments (typically time or
frequency).

The formal equivalence between the Schr\"odinger picture expansion on number states and the Heisenberg
picture representation of operators is useful for calculations in the few-mode but not in the cw case.
Both the Schr\"odinger and Heisenberg formalisms can be applied to predict mean values and variances,
but the quantum state reduction and dynamics following a detection event at time $t_c$, is not well
accounted for in the Heisenberg picture, and the dimensionality of the Hilbert space for infinitely
many modes makes the Schr\"odinger description unmanageable. We have previously shown that for
analyses restricted to Gaussian states and operations the full multi mode state reduction dynamics is
indeed tractable \cite{magnetometry,larsklaus,vivisq}. It is this latter approach that has been
extended in the present work, where we explicitly reduce the Gaussian field description of the full
multi-mode problem to a Gaussian field description of just two modes, from which the Schr\"odinger
representation of the state reduction mechanism is readily tractable, also in the non-Gaussian case.

The author acknowledges discussions with Jonas S. Neergaard-Nielsen, Eugene S. Polzik and Uffe V.
Poulsen.

\end{document}